     \documentstyle[12pt,epsfig]{article}
     \newlength{\dinwidth}                       
     \newlength{\dinmargin}                      
\def\lsim{\mathrel{\rlap{\lower4pt\hbox{\hskip1pt$\sim$}}
    \raise1pt\hbox{$<$}}}                
\def\gsim{\mathrel{\rlap{\lower4pt\hbox{\hskip1pt$\sim$}}
    \raise1pt\hbox{$>$}}}                
\newcommand{\GeV}{{\rm GeV}}
\newcommand{\MeV}{{\rm MeV}}

\begin{document}
\thispagestyle{empty}
\begin{flushleft}
DESY 99--072 \hfill
{\tt hep-ph/0008097}\\
August 2000  \\
\end{flushleft}

\setcounter{page}{0}

\mbox{}
\vspace*{\fill}
\begin{center}{\Large \bf
Leptoquark Pair Production in $\boldmath \gamma\gamma$
Scattering:\\  

\vspace{2mm}
Threshold Resummation$^{\normalsize
\footnotemark}$
\footnotetext{Contribution to the Proceedings of the 1998/99
Linear Collider Workshop.\\
Work supported in part by EU contract 
FMRX-CT98-0194(DG 12 - MIHT)}}

\vspace{5em}{
\large
Johannes Bl\"umlein$^a$ and Alexander Kryukov$^{a,b}$}
\\
\vspace{5em}
\normalsize
{\it $^a$Deutsches Elektronen-Synchrotron, DESY\\
Platanenallee 6, D-15738 Zeuthen, Germany}\\

\vspace{3mm}
{\it $^b$Institute of Nuclear Physics,\\
     Moscow State University, RU--119899 Moscow, Russia}\\

\vspace*{\fill}
\end{center}
\begin{abstract}
\noindent
The possibilities to pair--produce leptoquarks in photon--photon
collisions are discussed. QCD threshold corrections lead to a strong
enhancement of the production cross section. Suitably long--lived
leptoquarks ($\Gamma_\Phi \lsim 100 \MeV$) may form  Leptoquarkonium
states.
\end{abstract}

\vspace{1mm}
\noindent

\vspace*{\fill}
\newpage
\vspace*{10mm}
\begin{center}  \begin{Large} \begin{bf}
Leptoquark Pair Production in $\boldmath \gamma\gamma$
Scattering:\\
Threshold Resummation
\\
  \end{bf}  \end{Large}
  \vspace*{5mm}
  \begin{large}
Johannes  Bl\"umlein$^a$ and Alexander Kryukov$^b$ \\
  \end{large}

\vspace{2mm}
     $^a$Deutsches~Elektronen-Synchrotron~DESY,
     Platanenallee 6,~D-15738~Zeuthen,~Germany\\

\vspace{2mm}
     $^b$Institute of Nuclear Physics,\\
     Moscow State University, RU--119899 Moscow, Russia\\
\end{center}
%
\begin{quotation}
\noindent
{\bf Abstract:}
The possibilities to pair-produce leptoquarks in photon--photon
collisions are discussed. QCD threshold corrections lead to a strong
enhancement of the production cross section. Suitably long--lived
leptoquarks ($\Gamma_\Phi \lsim 100 \MeV$) may form  Leptoquarkonium
states.
\end{quotation}
%

\vspace{1mm}
\noindent
Leptoquarks are hypothetical particles which combine quantum numbers
of the fundamental fermions of the Standard Model and emerge as bosonic
(scalar and vector) states in various extensions of the Standard Model 
such as unified
theories and sub-structure models. In most of the scenarios the mass
spectrum of these states is not predicted. 
In a series of models, however,
one
expects states in the range of several hundred GeV to a few TeV. These
particles can be searched for at the next generation colliders as LHC
and future $e^+e^-$ linear colliders. Currently
the following mass ranges are
excluded by experiment for scalar leptoquarks:
\begin{eqnarray}
{\rm 1st~generation~leptoquarks~:}~~~~M &>& 242~\GeV~\cite{E1}~~
({\rm CDF~+~D0}) \nonumber \\
{\rm 2nd~generation~leptoquarks~:}~~~~M &>& 202~\GeV~\cite{E2}~~
({\rm CDF}) \nonumber \\
{\rm 2nd~generation~leptoquarks~:}~~~~M &>& 200~\GeV~\cite{E3}~~
({\rm D0}) \nonumber  \\
{\rm 3rd~generation~leptoquarks~:}~~~~M &>& ~99~\GeV~\cite{E4}~~
({\rm CDF}) \nonumber \\
{\rm 3rd~generation~leptoquarks~:}~~~~M &>& ~94~\GeV~\cite{E5}~~
({\rm D0}) \nonumber
\end{eqnarray}
at $95~\%$~CL
irrespective of the size of the fermion--leptoquark couplings, which is 
limited to very small 
values~\cite{H1, SD} for most of the leptoquark species.
Somewhat higher bounds are derived for vector leptoquark states,
depending on the size of their anomalous couplings to the 
gluon~\cite{BBK2}.
Unlike the fermionic couplings the
couplings of the leptoquarks to the gauge bosons of the Standard Model
are known,~cf.~\cite{BR1}, and are of the size of the standard gauge
couplings of the fermions. Due to the smallness of the fermionic
couplings leptoquark pair production processes at high energy 
colliders~\footnote{This applies to high energy
 $pp$~\cite{BBK2,BBK3,B2},
$ep$~\cite{BBK4,BBK2,B2},
$\gamma e$~\cite{BBK2},
$\gamma \gamma$~\cite{BB,BBK2}
and
 $e^+e^-$~\cite{BR1,BBK5,BB,BF} collisions.}
allow to perform a widely model independent search for these states.

At $e^+e^-$ high energy linear colliders the largest production cross
sections are obtained for $e^+e^-$ annihilation.
This process was studied in detail in Refs.~\cite{BR1,BBK5} both for
scalar and vector leptoquark states. The QED radiative corrections
to scalar leptoquark pair production were calculated at
$O(\alpha \log(s/m_e^2), O((\alpha \log(s/m_e^2))^2)$ both for initial
and final state radiation as well as the $O(\alpha_s)$ QCD correction
and the correction due to beamstrahlung in Ref.~\cite{B1}. 
These corrections are large in the threshold range. The enhancement due
to the QCD corrections, being dominated by the Coulomb singularity 
$\propto 1/\beta$ at low velocities is nearly balanced by the losses due
to the QED corrections and beamstrahlung despite of the smaller size
of the QED coupling constant. The second order QED corrections are still
of the size of $O(10\%)$ of the Born cross section and have therefore
to be taken into account.

A second important production channel is photon--photon pair production
of leptoquarks. Here we consider the case that the photon beams are
produced form the electron and positron--beams, respectively, by Laser 
beam Compton back--scattering. The photon
energy spectrum~\cite{NOVO} is described by
\begin{eqnarray}
\Phi_{\gamma}(z) &=& \frac{1}{N(x)}\left[1 - z + \frac{1}{1-z} -
\frac{4z}{x(1-z)} + \frac{4z^2}{x^2(1-z)^2}\right] \\
N(x) &=& \frac{16 + 32 x + 18 x^2 + x^3}{2x(1+x)^2} 
+ \frac{x^2-4x-8}{x^2} \log(1+x)~,
\end{eqnarray}
where $z$ denotes the longitudinal momentum fraction of the photons
after beam conversion and $x = 2(\sqrt{2}+1)$, with $z \leq x/(1+x)$.
Alternatively, one may consider leptoquark pair production by
photon--photon scattering preparing the initial state through
Weizs\"acker--Williams emission from the $e^+e^-$ beams.
These contributions are, however,
much smaller than those due to Compton--conversion, cf.~\cite{BB,
BBK2}.
The photon--photon cross section reads
\begin{eqnarray}
\sigma_{\Phi  \overline{\Phi}  }(s) =  \int_0^{z_{max}} dz_1
                                       \int_0^{z_{max}} dz_2
\Phi_{\gamma}(z_1) \Phi_{\gamma}(z_2)~
\hat{\sigma}_{\Phi  \overline{\Phi}  }(z_1 z_2 s)~\theta(z_1 z_2 s
- 4 M_{\Phi}^2)~.
\end{eqnarray}
For scalar leptoquarks the direct contribution to the sub--system
cross section is given by
\begin{eqnarray}
\hat{\sigma}_{\Phi_S\overline{\Phi}_S}(s) =
\frac{\pi \alpha^2}{s} Q_{\Phi}^4 \left[ 2(2-\beta^2) \beta - (1-\beta^4)
\log \left|\frac{1+\beta}{1-\beta}\right|\right]~.
\end{eqnarray}
The corresponding relations for vector leptoquarks were derived
in~\cite{BB} and are somewhat lengthly due to the emergence of
anomalous couplings.

Besides the direct contributions $\gamma \gamma \rightarrow
\Phi \overline{\Phi}$ direct--resolved and resolved--resolved
terms are present,
\begin{eqnarray}
\sigma(\gamma + \gamma \rightarrow \Phi \overline{\Phi}) 
= \sigma_{dir.}
                                         + \sigma_{res.,dir.}
                                         + \sigma_{res.}~
\end{eqnarray}
The latter ones are hadronic contributions
and were calculated for both scalar and vector leptoquarks in 
Refs.~\cite{BBK4,BB,BBK2} including two anomalous couplings for
vector leptoquarks. The numerical analysis shows, that these terms
contribute significantly only far away from threshold, i.e. typically
for
larger values of the cms velocities $\beta \geq 0.8$ of the 
leptoquarks~\cite{BBK2}.

In the search region (threshold) the photo--pair production cross
sections behave like
   $\propto Q_{\Phi}^4$ and may
vary by a factor of 625 between the production cross sections
for $|Q_{\Phi}| =1/3$ and for $|Q_{\Phi}| =5/3$ states, which have the
same cross section at a hadron collider.  For     vector leptoquark
the cross sections are strongly sensitive to the anomalous couplings
$\kappa_\gamma$ and $\lambda_\gamma$~\cite{BB,BBK2}.

In the threshold region QCD corrections to the photon--photon process,
similar as in $e^+e^-$ annihilation~\cite{B1}, are very important.
These corrections can only be calculated for the case of scalar
leptoquark pair production, since for vector leptoquarks the effective
Lagrangian does not correspond to a renormalizable theory. We will
therefore limit the consideration to the case of scalar leptoquarks
here. Threshold resummations of the universal terms have been considered
in the literature before, cf. Refs.~\cite{FAD,FBK,SP}. For a
final state of a pair of scalar leptoquarks we follow \cite{Bigi}.
The Born cross section $d\sigma_B$ obtains the correction factor
$K_S(E)/K_S^{(B)}(E)$,
\begin{eqnarray}
d \sigma = d \sigma^{\rm BORN} \frac{K_s(E)}{K_S^B(E)}
\end{eqnarray}
\begin{eqnarray}
K^{(B)}(E) &=& \frac{M_{\Phi}}{4\pi} \sqrt{M_{\phi} E} \\
K_S(E)     &=& \frac{M_{\Phi}^2}{4\pi} 
\left\{ \frac{k_+}{M_{\Phi}} 
+ \frac{2 k_1}{M_{\Phi}} 
{\rm arctan} \left(\frac{k_+}{k_-}\right) \right.
\nonumber\\  & &  \left.
~~~~~~~~~~~~~~+ \sum_{n=1}^{\infty} \frac{2k_1^2}{M_{\Phi}^2 n^4}
\frac{
\Gamma_{\Phi} k_1 n 
+ k_+ \left[n^2 \sqrt{E^2+\Gamma_{\Phi}^2} + k_1^2/M_{\Phi}\right]}
{\left[E+k_1^2/(M_{\Phi} n^2)\right]^2 + \Gamma_{\Phi}^2} \right\}~,
\end{eqnarray}
with
$E =  \sqrt{s} - 2 M_{\Phi}$. $M_S$ and $\Gamma_S$ denote the mass
and width of the scalar leptoquark and
\begin{eqnarray}
k_1 &=& \frac{2}{3} \alpha_s(M_{\Phi}) M_{\Phi} \nonumber\\
k_{\pm} &=& \sqrt{\frac{M_{\Phi}}{2} \left(\sqrt{E^2 + \Gamma_{\Phi}^2}
\pm E\right)}~. \nonumber
\end{eqnarray}
Here the strong coupling constant $\alpha_s = \alpha_s(M^2_S)$
was considered to be fixed.

In Figure~1 the ratio of the integrated cross section with threshold
resummation and the Born cross section is shown assuming a decay width of
$\Gamma_S = 1 \GeV$ as an illustration.~\footnote{The 'idealized' decay 
widths are by
far smaller due to the the small fermionic couplings. However, the
real leptoquark width should be larger since these particles are
supposed to fragment into hadrons similar as quarks do. A theoretically
safe value of the width is hard to obtain due to these non--perturbative
effects.} Threshold enhancements of a factor 5--8 can be obtained.

If the width of the leptoquarks turns out to be $\lsim 100~\MeV$
{\it Leptoquarkonia} can be formed in $\gamma \gamma$--collisions, cf.
~\cite{B1}~\footnote{This
possibility was later discussed in \cite{KIS,BC} too.}.
The $\beta$--behavior at threshold is favoring
the $\gamma \gamma$--process ($\sigma
\propto \beta$) in comparison to the
$e^+e^-$ process ($\sigma \propto \beta^3$).
In Figure~2 the total cross sections are shown with and without
threshold resummation.

To summarize, photon pair--production of scalar and vector leptoquarks
at future $e^+e^-$ linear colliders were studied. The cross sections
vary by the leptoquark charge as $|Q_\Phi|^4$, which may mean a variation
up to a factor of 625. This production process offers a background free
window to study the anomalous couplings $\kappa_A$ and $\lambda_A$
of potential vector leptoquark
states to the photon. The threshold QCD enhancement of the 
photon--photon process is of $O(5-8)$ for typical choices of the
parameters. Leptoquarkonia can be formed in the photon--photon
process iff their width is $\Gamma_\Phi \lsim 100 \MeV$.



\newpage
\begin{center}

\mbox{\epsfig{file=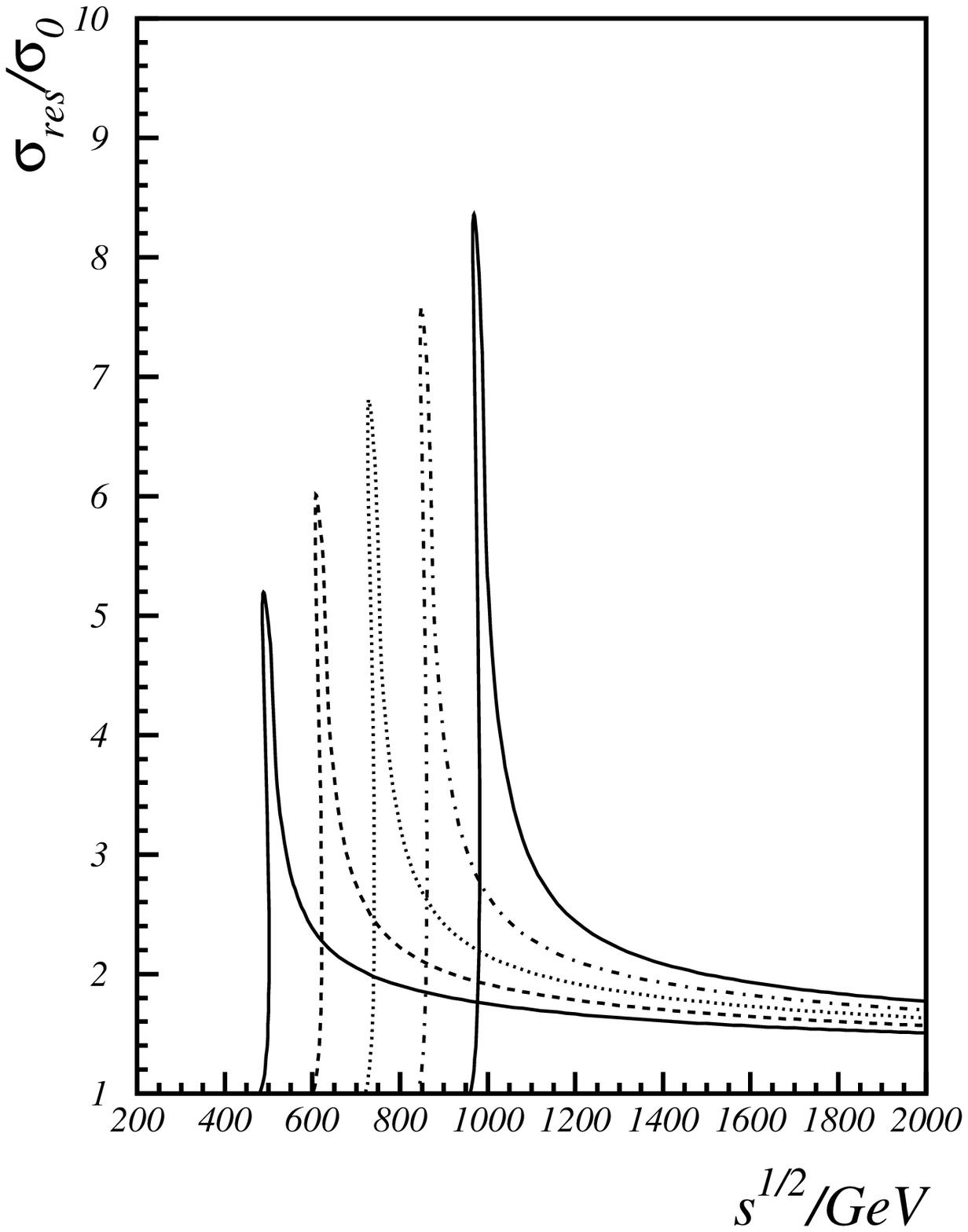,height=18cm,width=16cm}}

\vspace{2mm}
\noindent
\small
\end{center}
{\sf
Figure~1:~Ratio of the leptoquark pair production cross section including
the threshold resummation and the Born cross section as a function of 
$\sqrt{s}$.
Left full line: $M = 200~\GeV$; dashed line:
$M = 250~\GeV$; dotted line: $M = 300~\GeV$, dash--dotted line:
$M = 350~\GeV$; right full line: $M=400~\GeV$.}
\normalsize
\newpage
\begin{center}

\mbox{\epsfig{file=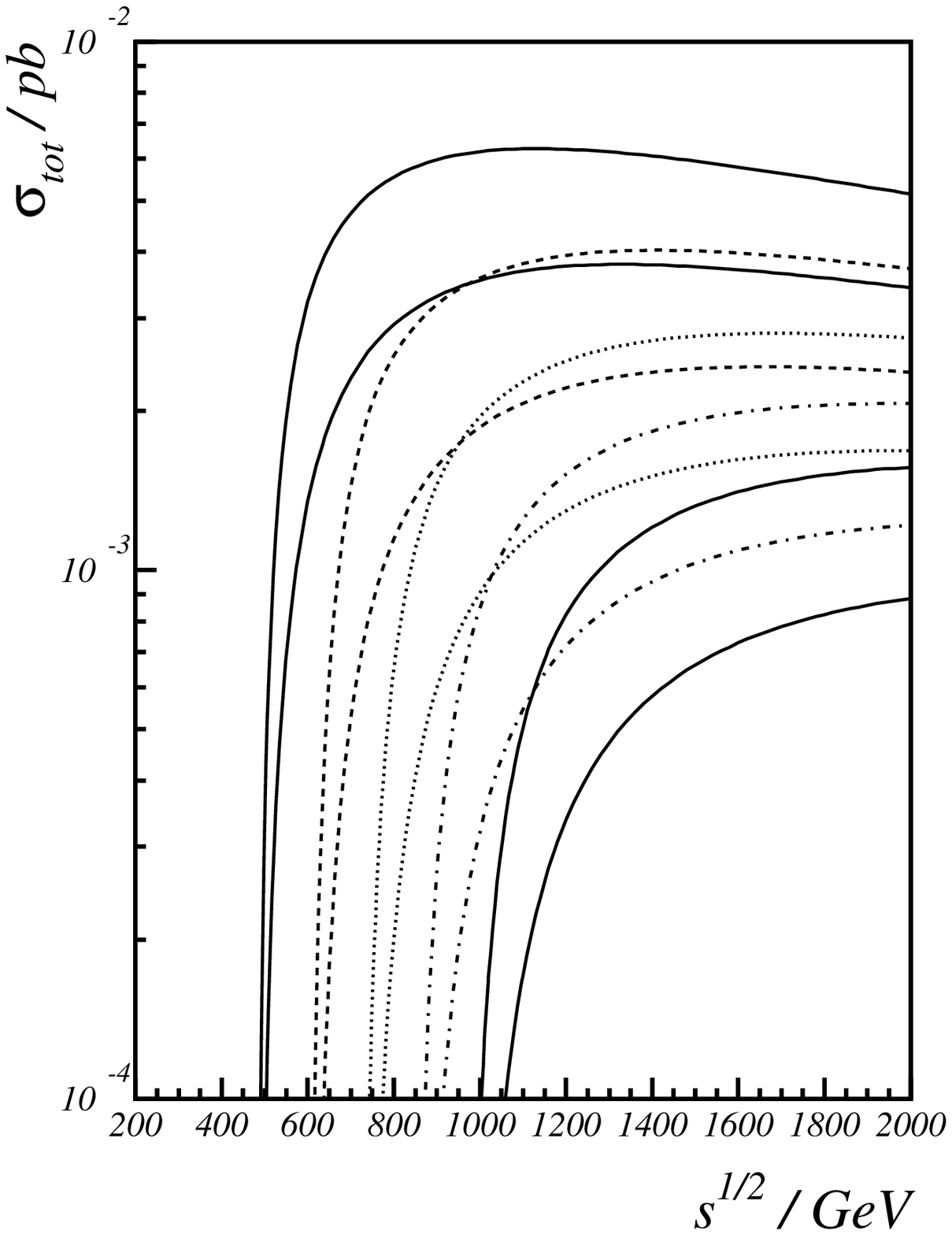,height=18cm,width=16cm}}

\vspace{2mm}
\noindent
\small
\end{center}
{\sf
Figure~2:~Leptoquark pair production cross section $\sigma(\gamma \gamma
\rightarrow \overline{\Phi} \Phi)$ as a function of $\sqrt{s}$.
Each pair of equally drawn lines corresponds to the same leptoquark mass.
Upper line: $\sigma^{\rm Born} + \sigma_{resum}$; lower line:
$\sigma^{\rm Born}$. Upper full lines: $M = 200~\GeV$; dashed lines:
$M = 250~\GeV$; dotted lines: $M = 300~\GeV$, dash--dotted lines:
$M = 350~\GeV$; lower full lines: $M=400~\GeV$.}
\normalsize
\end{document}